\input psfig
\input mn

%
\def\spose#1{\hbox to 0pt{#1\hss}}
\def\lta{\mathrel{\spose{\lower 3pt\hbox{$\mathchar"218$}}
     \raise 2.0pt\hbox{$\mathchar"13C$}}}
\def\gta{\mathrel{\spose{\lower 3pt\hbox{$\mathchar"218$}}
     \raise 2.0pt\hbox{$\mathchar"13E$}}}
%



\loadboldmathnames



\pagerange{}    
\pubyear{1997}
\volume{}

\begintopmatter  

\title{A model for WZ Sge with ``standard" values of $\alpha$}

\author{J.-M. Hameury$^1$, J.-P. Lasota$^2$ and J.-M. Hur\'e$^3$}

\affiliation{$^1$ URA 2180 du CNRS, Observatoire de Strasbourg, 11 rue de 
l'Universit\'e, F-67000 Strasbourg, France}
\smallskip
 \affiliation{$^2$ UPR 176 du CNRS, DARC, Observatoire de Paris, Section de
Meudon, F-92195 Meudon Cedex, France}
\smallskip 
\affiliation{$^3$ URA 173 du CNRS, DAEC, Observatoire de Paris, Section de
Meudon, F-92195 Meudon Cedex, France}
\smallskip
\shortauthor{J.-M. Hameury, J.-P. Lasota and J.-M. Hur\'e}
\shorttitle{A model for WZ Sge with ``standard" values of $\alpha$}



\abstract {We present a model for the dwarf nova WZ Sge which does not
require assuming unusually and unexplained low values of the viscosity
$\alpha$ -- parameter during exceptionally long quiescent states of
this system.  We propose that the inner parts of the accretion disc are
disrupted by either a magnetic field or evaporation, so that the disc
is stable (or very close to being stable) in quiescence, as the mass
transfer rate is very low and the disc can sit on the cool, lower
branch of the thermal equilibrium curve.  Outbursts are triggered by an
enhanced  mass transfer, which brings the disc into the unstable regime
of the standard dwarf nova disc instability model. The resulting
eruptions are strongly affected by the irradiation of the secondary
star. Our model reproduces very well the recurrence time and the
characteristics of the light curve in outburst.  }


\maketitle  

\section{Introduction}

Dwarf novae (DN) are cataclysmic variables, which, at usually irregular
intervals, undergo eruptions in which the brightness increases by 2 to
7 magnitudes. It now well established that dwarf nova eruptions have
their origin in a local, thermal and viscous instability due to an
abrupt change in opacities at densities and temperatures at which
hydrogen is partially ionized. In the disc instability model (DIM) of
dwarf nova outbursts one assumes that the $\alpha$ viscosity parameter
is higher in the high state than in quiescence and on obtains in this
way a global disc instability which gives a rather good description of
the U~Gem type dwarf nova properties. Current models of dwarf novae
imply values of the $\alpha$ parameter of the order of 0.01 in the low,
quiescent state, and about 4 -- 10 times larger during outbursts.

The DIM in its standard version cannot however describe outbursts of Z~Cam
and SU~UMa type of dwarf novae.  In the `standard' DIM  one assumes
that the mass transfer from the secondary star is constant, that the
only instability operating in the disc is the thermal--viscous one, and one
neglects illumination of the disc and of the secondary by the radiation
emitted by the accretion on to the white dwarf during outbursts.
It is clear that if the Z~Cam and SU~UMa type of dwarf nova eruptions
are due to the same type of local instability as the one that operates
in U~Gem type dwarf nova systems at least one of the standard assumptions
must be dropped.

In the case of SU~UMa type dwarf novae it has been proposed by Osaki
(see e.g. 1995) that their superoutbusts are due to a tidal-thermal
instability (TTI). In this model, successive U~Gem type outbursts
(`normal' outbursts) lead to an accumulation of matter in the disc and
an increase of its outer radius. When the disc's outer rim enters the `tidal
radius' in which the 3:1 resonance operates, the tidal couple is
assumed to increase the accretion rate and to trigger a superoutburst
in which matter accumulated during the supercycle is dropped on to
the white dwarf. The TTI model encounters several difficulties. One is
that it cannot apply (see e.g Smak 1996) to the 1985 superoutburst
(Mason et al. 1988) of U~Gem since the tidal instability cannot operate
in this system.  Another one is connected with the superoutbursts of
WZ~Sge. The TTI model can describe this system, which shows only
superoutbursts separated by very long quiescent intervals, only if one
assumes a very low value of $\alpha \lta 10^{-4}$ (Smak 1993; Osaki
1995). The reason for such a low $\alpha$ in this particular system is
however unexplained so that it suggests that the TTI model might not be
the correct description of the WZ~Sge behaviour.

Lasota et al. (1995) noticed that if the inner disc in WZ Sge were missing
because of the presence of a magnetosphere (Livio \& Pringle 1992) or
because of evaporation (Meyer \& Meyer--Hofmeister 1994) the accretion
disc would be marginally stable and outbursts could be triggered by
an enhanced mass transfer (EMT) from the secondary. The long recurrence
time would then be the timescale of fluctuations of the mass transfer
and the $\alpha$ parameter would have its `usual' value.
The idea that superoutbursts in general could be due to EMT has been
proposed by Smak (see e.g. Smak 1996).

Warner et al. (1996) assert that in the case of a truncated inner disc
one can obtain long recurrence times in the `standard' DIM with a
`standard' value of the $\alpha$ parameter. As we shall show below, the
Warner et al.  (1996) model is is fact, as far as recurrence time is
concerned, practically identical to the one proposed by Lasota et al.
(1995). The `standard' DIM with `standard' values of $\alpha$ suffers
however from one insurmountable difficulty: the mass contained in the
quiescent accretion disc for the parameters describing WZ Sge and for
$\alpha \sim 0.01$ is less than $10^{23}$ g whereas during the
superoutburst of this dwarf nova more than $10^{24}$ g has been
accreted by the central white dwarf (Smak 1993). In fact this
difficulty is the main reason for invoking very low values of the
viscosity parameter.

If one wishes to explain superoutbursts of WZ Sge without using 
extremely low values of $\alpha$ one cannot avoid adding mass prior to
or during the eruption. As discussed by Smak (1996) there is evidence
of an increased mass transfer rate during normal outbursts and superoutbursts.
In the EMT scenario proposed by Smak (1996) the superoutburst begins with
a major enhancement of the mass transfer caused by irradiation during
the preceding normal outburst. In the case of WZ Sge which shows only
superoutbursts such a trigger does not exist. 

In the present article we show that if, as proposed by Lasota et al.
(1995), the quiescent accretion disc in WZ Sge is marginally stable, a
slight enhancement of mass transfer triggers a `normal' outburst which,
due to subsequent irradiation of the secondary, becomes a superoutburst
since a substantial amount of matter is added to the disc during 
the active phase of the cycle.

\section{WZ Sge in quiescence}

The orbital period of WZ Sge is 81 min, close to the minimum period of
cataclysmic variables; the system is thus very compact, and the
accretion disc is quite small, with an outer radius of 1.1 10$^{10}$ cm
during quiescence (Smak 1993). Here and in what follows we shall use
Smak's (1993) model of WZ Sge according to which the primary and
secondary masses are 0.45 and 0.06 M$_\odot$ respectively. From the
luminosity of the hot spot, Smak (1993) determined a mass transfer rate
in quiescence $\dot{M}_{\rm tr} \approx 2 \ 10^{15}$ g $s^{-1}$.

WZ Sge has been detected by Einstein (Eracleous et al. 1991), EXOSAT
(Mukai \& Shiokawa 1993) and ROSAT (van Teeseling et al. 1996); its
X-ray luminosity did not significantly vary in the interval  from 4
months to more than 7 years after the 1978 outburst, and stayed at the
level of 3 10$^{30}$ erg s$^{-1}$. If interpreted in terms of accretion
rate on to the white dwarf surface, this luminosity corresponds to an
accretion rate $\dot{M}_{\rm acc}$ on to the white dwarf surface of:
$$
\dot{M}_{\rm acc} = 5.0 \times 10^{13} \eta_X^{-1} r_9 \left( {M_1 \over 0.45
\rm M_\odot} \right)^{-1} \left( {L_{\rm X} \over 3 \times 10^{30}} \right)
\; \rm g s^{-1} \eqno \stepeq
$$
where $M_1$ is the primary mass, $r_9$ the radius in units of 10$^9$ cm,
$L_{\rm X}$ the X-ray luminosity and $\eta_X \leq 1$ the efficiency conversion
of gravitational energy into X-ray photons.  This would imply that the disc
is not very far from steady state, since the mass accretion rate did not vary
significantly from soon after the outburst to now, i.e.  during an interval
which is comparable to the recurrence time.  

Moreover, according to the DIM,
in the non--equilibrium disc $\dot{M}_{\rm acc}$ has to be low enough so that
the inner regions of the disc are on the cool, stable branch of the $\Sigma -
T_{\rm ef\!f}$ curve, which implies: 
$$
\dot{M} <
\dot{M}_{\rm B} = 1.43 \times 10^{13} M_1^{-0.87} r_9^{2.60} \; \rm g
s^{-1} \eqno \stepeq 
$$ 
(Ludwig et al. 1994). Since the accretion
disc has to be in this state very early after outburst one concludes, 
comparing Eqs. (1) and (2), that in
the DIM framework the quiescent viscosity cannot be very small, i.e. that
$\alpha$ cannot be very small.

Consistency between X-ray observations and Eq. (2) requires the radius
of the inner edge of the disc $r_{\rm in}$ to be larger than:
$$
r_{\rm in} > 1.43 \times 10^{9} \eta_X^{-0.62} \left( {M_1 \over 0.45 \rm
M_\odot} \right)^{0.08} \; \rm cm \eqno \stepeq
$$
This is larger than the white dwarf radius, which is easily understood if the
inner parts of the disc are disrupted by either a magnetic field (see e.g. 
Livio \& Pringle 1992), or by evaporation (Meyer \& Meyer-Hofmeister 1994). 
At such radii however, the disc is very close to a globally stable
configuration (Lasota et al. 1995; Warner et al. 1996).  In the following, we
assume $r_{\rm in} = 4 \times 10^9$ cm, which ensures that the total disc
luminosity is less than the luminosity of the hot spot, in agreement with
observations.

\section{The outburst}

As mentioned in the introduction two ``standard $\alpha$"  mechanisms have
been proposed for triggering outbursts in WZ~Sge. In one of them, the disc is
supposed to be stable on the cool branch in quiescence, in which case
outbursts have to be triggered by an increase of the mass transfer from the
donor star, causing the disc to become thermally unstable (Lasota et al.
1995); in the other one, the disc is supposed to be  marginally unstable
and to undergo a `standard' DIM outburst.
(Warner et al. 1996).  It must however be noted that, because the disc must
remain on the cool, stable branch during quiescence, its mass cannot be more
than the integral of the maximum surface density $\Sigma$, i.e. $M_{\rm d} <
M_{\rm max} = 6 \times 10^{21} \alpha^{-0.8}$ g (Smak 1993). If the mass
transfer rate from the secondary remains constant, the recurrence time can be
estimated as the time it takes to increase $M_{\rm d}$ to this maximum value,
i.e.
$$
\eqalignno{
t_{\rm rec} & \sim {M_{\rm max} \over \dot{M}_{\rm tr} - \dot{M}_{\rm acc}} \cr
& \sim 4 \left( {\alpha \over 10^{-2}} \right)^{-0.8} \left( 1 -
{\dot{M}_{\rm acc} \over \dot{M}_{\rm tr}} \right)^{-1} \; \rm yr & \stepeq\cr}
$$
If one requires that $\alpha$ is not extremely small, $\dot{M}_{\rm acc}$
must be very close to $\dot{M}_{\rm tr}$ (within 10\%) during the whole
quiescent phase in order to get very long recurrence times.  It is therefore
most likely that an outburst would be triggered by a small fluctuations of
$\dot{M}_{\rm tr}$ that is expected to occur within the long recurrence time
of WZ Sge. Since in the Lasota~et~al.~(1995) model the disc is marginally
stable, both models are in fact identical as far as the recurrence time is 
concerned. 

In the standard DIM, the duration of the outburst is at most the time
it takes to empty the disc, i.e.
$$ \eqalignno{ t_{\rm outb} & =
{M_{\rm max} \over \dot{M}_{\rm acc}-\dot{M}_{\rm tr}} \cr & \sim 3
\left( {\alpha \over 10^{-2}} \right)^{-0.8} \left( 1 - {\dot{M}_{\rm
tr} \over \dot{M}_{\rm acc}} \right)^{-1}  \; \rm days & \stepeq\cr} 
$$
where we took $\dot{M}_{\rm acc} = 10^{18}$ g s$^{-1}$ (Smak 1993).  The long
duration of WZ Sge outbursts (about one month) implies, if $\alpha$ is not
very small, that, during outburst, the mass transfer rate had increased by
two orders of magnitude, and was of the same order as the accretion rate. 
This would very naturally result from the illumination of the secondary.
Effects of irradiation of the secondary were clearly observed during the
outburst of SS Cyg (Hessman et al. 1984) and an
increase of $\dot{M}_{\rm tr}$ by factors $\sim$ 2 has been observed in
DNs such as Z Cha and U Gem (Smak 1995).  Illumination
effects are expected  to be even more important in WZ Sge, in which the
quiescent $\dot{M}_{\rm tr}$ is particularly low, so that the secondary
surface temperature is expected to be low, and which has the shortest orbital
period, so that the X-ray flux heating the secondary is large. For example,
Smak~(1993) obtains for the effective temperature of the secondary during the
outburst $T_{\rm eff, 2} \approx 17 000$ K.

In the simplest model for
illumination, the mass transfer rate is proportional to $e^{-\Delta r / H}$,
where $\Delta r$ is the distance between the secondary photosphere and the
Lagrangian point $L_1$, and $H$ the atmospheric scale height, proportional to the
secondary surface temperature. In the case of WZ Sge, $\Delta r / H$ would
typically be $\gta 1$ in quiescence, and less than unity during outburst,
leading to large variations of the mass transfer rate.  
The response of the secondary to illumination has been discussed by
e.g. Osaki (1985) and Hameury~et~al.~(1986), but the presence of
screening effects, of flows from the secondary's poles to $L_1$, and the
dependence on the emitted spectrum are very complex and make it
difficult to describe; the exponential dependence quoted above is certainly
not a very good approximation. As a preliminary step, we assume here that
$$
\dot{M}_{\rm tr} = \gamma \dot{M}_{\rm acc}
\eqno \stepeq
$$
with $\gamma < 1$; this is similar to the approach of Augusteijn~et~al.~
(1993) in the context of soft X--ray transients.

\beginfigure{1}
\psfig{file=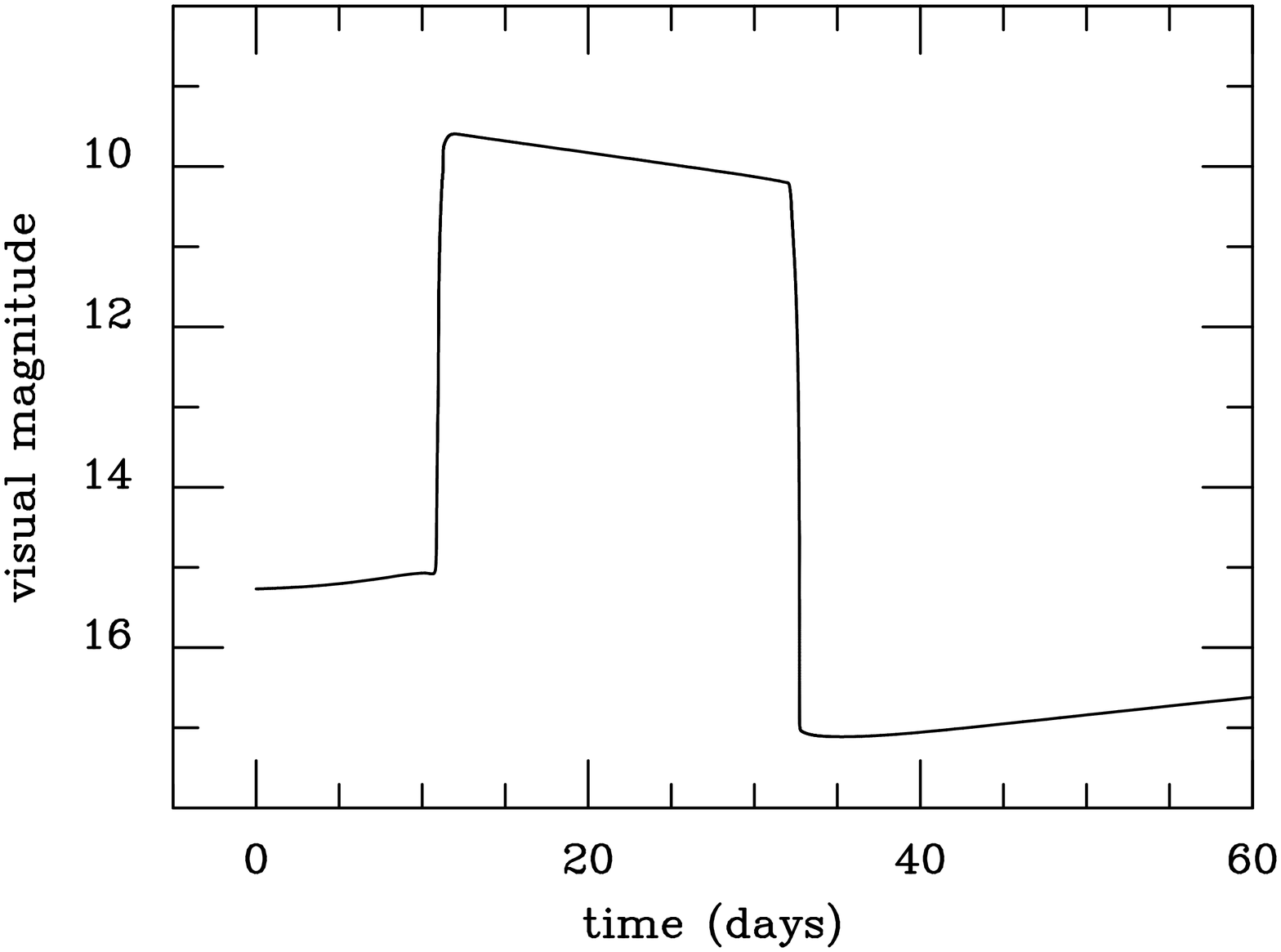,width=\ColumnWidth}
\caption{{\bf figure 1} Predicted visual light curve for the outburst 
of WZ Sge}
\endfigure

The expected outcome of the model is that during quiescence, the disc stays
on the cool stable branch. A fluctuation of the mass transfer rate from the
secondary triggers the viscous/thermal instability, most probably at the
outer edge of the disc. A heat front then propagates towards the compact
object.  Once it reaches the inner edge of the disc, $\dot{M}_{\rm tr}$
increases up to a value $\gamma \dot{M}_{\rm acc}$, and then, since the
viscous time of the disc is short as compared to the total duration of the
outburst, the disc would be close to steady state, with a mass transfer rate
equal to $\gamma$ times the mass accretion rate, while the disc 
introduces a delay
equal to the viscous time $t_{\rm visc}$. This naturally produces an
exponential behaviour, with a decay time equal to $t_{\rm visc} /\ln \gamma$.
Eventually, $\dot{M}_{\rm tr}$ becomes less than the critical value below
which the hot, stable solution in the $\Sigma - T_{\rm ef\!f}$ diagram does
not exist any longer; then a cooling wave starts from the outer edge of the
disc, and brings it into quiescence in a short time scale.

In order to test this, we have calculated the time-dependent evolution of a
disc initially stable with a low  $\dot{M}_{\rm tr}$, taken to be $1.5 \times
10^{15}$ g s$^{-1}$, which is suddenly increased to $\max (5 \times 10^{15},\
0.87 \dot{M}_{\rm acc})$ g s$^{-1}$. 
The enhanced mass transfer decays exponentially in 5 days.
$\alpha$ is taken to be $0.01$ in the
cool branch and 0.1 in the hot one. All other parameters are those quoted
here for WZ Sge; the code used is described in Hameury et al. (1996). The
disc is assumed to be truncated as a result of the presence of a magnetic
field, so that:
$$
r_{\rm in} = 4 \times 10^9 \left( {\dot{M}_{\rm acc} \over 1.5 \times 10^{15}
\rm g s^{-1}} \right)^{2/7} \; \rm cm
\eqno \stepeq
$$
The disc behaves exactly as described above; for illustration, the visual
magnitude of the disc (i.e. that does not include contributions from the
secondary or from the white dwarf) is displayed in Fig. 1. It is seen that
the shape of the light curve is in good agreement with observations.

\section{Conclusion}

We have shown that the unusually long recurrence time and outburst duration
in WZ Sge does not require the viscosity in this system to be much lower than
in all other systems; these characteristics would result from (1) a low value
of the mass transfer rate, so that the system is marginally stable during
quiescence; (2) a truncated disc, which is required in many other systems to
account for e.g. the observed optical-UV delay; and (3) a significant
illumination effect that increases the mass transfer rate from the secondary
by two orders of magnitude. WZ Sge would thus be explained by a combination
of the two different model proposed by Osaki (1974, 1985), in a way similar
to the proposition of Duschl \& Livio (1989): a fluctuation of the mass
transfer rate produces a thermal/viscous disc instability that brings the
disc into a hot state, leading to an sudden increase of $\dot{M}_{\rm tr}$,
which then slowly decreases until a cooling wave rapidly brings the disc back
into its quiescent cool state.

The possible tests of this model are not very different from those proposed
by Warner et al. (1996), since the outburst is an outside-in outburst, and
since the enhancement of mass transfer that triggers the outburst is not very
large, so that the behaviour of the disc outer radius need not be very
different from that of a pure disc instability. We do however predict an
increase of the hot spot luminosity a few days before the onset of an
outburst; we also predict that the occurrence of outburst should be irregular
(for example of a shot noise type). Finally, the hot spot should be much
brighter during outbursts; this has been observed by Patterson et al. (1975)
who inferred an enhancement of mass transfer from the secondary by a factor
of 60 to 1000; they deduced then that the cause of the outburst was a mass
transfer instability, although they could not exclude the possibility ``that
a brightening of the white dwarf or disc could be the event that triggers
unstable mass transfer from the secondary".

\bye